\documentclass[fleqn,usenatbib]{mnras}
\usepackage{amssymb}

\usepackage{graphicx}                   
\usepackage{float}
\usepackage{amsmath}
\usepackage{amssymb}
\usepackage{multirow}
\usepackage{bm}
\usepackage{color}
\usepackage{ulem}

\usepackage{txfonts}

\usepackage[T1]{fontenc}

\DeclareRobustCommand{\VAN}[3]{#2}
\let\VANthebibliography\thebibliography
\def\thebibliography{\DeclareRobustCommand{\VAN}[3]{##3}\VANthebibliography}

\usepackage{graphicx}   
\usepackage{amsmath}    
\usepackage{amssymb}    

\title[The glitch recovery in a strangeon star model]{Pulsar Glitch in a Strangeon Star Model. III. The recovery}

\author[X. Y. Lai et al.]{
X. Y. Lai$^{1,2}$\thanks{E-mail: laixy@hue.edu.cn}, W. H. Wang$^{3}$, J. P. Yuan$^{4}$, R. P. Lu$^{5}$, H. Yue$^{5}$ and R. X. Xu$^{6,7}$
\\
$^{1}$Department of Physics and Astronomy, Hubei University of Education, Wuhan 430205, China\\
$^{2}$Research Center for Astronomy, Hubei University of Education, Wuhan 430205, China\\
$^{3}$College of Mathematics and Physics, Wenzhou University, Wenzhou 325035, China\\
$^{4}$Xinjiang Astronomical Observatory, Chinese Academy of
Sciences, Urumqi, XinJiang 830011, China\\
$^{5}$School of Earth and Space Sciences, Peking University, Beijing 100871, China\\
$^{6}$School of Physics, Peking University, Beijing 100871, China\\
$^{7}$Kavli Institute for Astronomy and Astrophysics, Peking
University, Beijing 100871, China}

\date{Accepted XXX. Received YYY; in original form ZZZ}

\pubyear{2021}

\begin{document}
\label{firstpage}
\pagerange{\pageref{firstpage}--\pageref{lastpage}}
\maketitle

\begin{abstract}
Strangeon star model has passed various observational tests, such as
the massive pulsars and the tidal deformability during binary
mergers.
Pulsar glitch, as a useful probe for studying the interior structure
of pulsars, has also been studied in strangeon star model in our
previous papers, including the recovery coefficient, the waiting
time of glitches and glitch activity.
In this paper, the recovery process of a glitch is described in the
strangeon star model, based on the starquake picture established
before (in Paper I).
After the starquake, the inner motion of the stellar matter would
reduce the tangential pressure in the cracked places at the
equatorial plane.
The recovery (increase) of the tangential pressure would be achieved
by a viscous flow towards the cracked places at equatorial plane,
which leads to the exponential recovery of the spin frequency.
A uniform viscous flow can reproduce the single exponential decay observed in some glitches, and the viscous time-scale $\tau$ and the depth $h$ of the cracking place below the surface can be fitted by the recovery data.
It is found that $h$ increases with glitch size $\Delta\nu/\nu$, which is expected in the glitch scenario of strangeon stars.
The magnitude of the recovery predicted in this recovery model is also consistent with that derived from observations.
The single exponential decay reproduced by a uniform viscous flow can be generalized to two or more exponentials by the multi-component of viscous flows.
\end{abstract}




\section{Introduction}

The theoretical difficulty in solving the non-perturbative quantum
chromodynamics (QCD) problems, however, makes it challenging to
describe the state of supranuclear matter in pulsar-like compact
stars.
The perturbative QCD, based on asymptotic freedom, works well only at high energy scales, $E_{\rm scale}>\Lambda_{\rm QCD} \sim 1$ GeV.
However, the state of pressure-free strong matter at supra-nuclear density should be relevant to non-perturbative QCD because $E_{\rm scale} < \Lambda_{\rm QCD}$, which is exactly a similar case of normal atomic nuclei.
Starting from deconfined quark state with the inclusion of strong interaction between quarks, and using the Dyson-Schwinger-Equation approach to the non-perturbative QCD~\citep{Fischer2006}, one would estimate that the strong coupling constant $\alpha_{\rm s}$ could be greater than 1 at density $\sim 3 \rho_0$.
It is worth noting that a weakly coupling strength comparable with that of quantum electrodynamics is possible only if the baryon number density is much larger than of a pulsar's core region, so a weakly coupling treatment is inadequate for realistic dense matter in pulsars.
From this point of view, although some efforts have been made to understand the state of pulsar-like compact stars in the framework of conventional quark stars, including the MIT bag model with almost free quarks~\citep{Alcock1986} and the color-superconductivity state model~\citep{Alford2008}, realistic densities inside pulsar-like compact stars cannot be high enough to justify the validity of perturbative QCD.

The strong coupling between quarks may render quarks grouped in quark-clusters, and each quark-cluster/strangeon is composed of several quarks condensating in position space rather than in momentum space.
A conjecture of ``condensation'' in position space, with strange quark cluster as the constituent units~\citep{Xu2003}, rather than condensation in momentum space for a color super-conducting state, was thus made for cold matter at supra-nuclear density.
The strange quark cluster is renamed strangeon, being coined by combining ``strange nucleon'' for the sake of simplicity~\citep{XG2016,Wang2017ApJ}.
Strangeon matter is conjectured to be the compressed baryonic matter of compact stars, where strangeons form due to both the strong and weak interactions and become the dominant components inside those stars.
Although whether quarks inside pulsars would be grouped in strangeons is hard to answer from direct QCD calculations, the astrophysical point of view could give us some hints.

Compact stars composed totally of strangeons are called ``strongeon stars''.
It is proposed that the pulsar-like compact stars could actually be strangeon stars, and this proposal is also motivated by several astrophysical points of view~\citep{Xu2003}.
Being similar to strange quark stars, strangeon stars have almost the same composition from the center to the surface.
The properties of strangeons, as well as the strangeon matter surface, could then be calculated, to be helpful for understanding different observations of pulsar-like compact stars (see the reviews~\cite{LX2017,LXX2023} and references therein).
For example, strangeon star model predicts high mass pulsars~\citep{LX2009a,LX2009b} before the discovery of pulsars with $M>2M_\odot$~\citep{Demorest2010}, and the tidal deformability~\citep{LZX2019} as well as the light curve~\citep{Lai2018RAA,Lai2021RAA} of merging binary strangeon stars are consistent with the results of gravitational wave event GW170817~\citep{GW170817} and its multiwavelength electromagnetic
counterparts~\citep[e.g.,][]{Kasliwal2017,Kasen2017}.

Strangeon stars would have global rigidity, since the mass of a strangeon that comprised of a large number of quarks is large and the formation of solid structures become likely due to the short quantum wavelength $\sim h/(mc)$.
The melting temperature would be much larger than the temperature inside pulsars~\citep{Xu2003}.
Moreover, because of their classical behavior (as a large mass may result in a small quantum wavelength), strangeons that exist in compact stars could locate in periodic lattices (i.e., in a solid state) when temperature becomes sufficiently low ($kT<10$ MeV).
The global rigidity would have broad astrophysical interests and have significant implications in astronomical observations.
Starquakes of solid strangeon stars could induce glitches.
In this paper, we will discuss the glitch recovery of strangeon
stars, in the framework of starquakes.

A pulsar glitch is an impulsive spin-ups followed by a
quasi-exponential recovery towards the steady spin-down.
This kind of timing irregularity is widely accepted as a window into
the interior structure of pulsar-like compact stars.
The nature of pulsars, in turn, is the starting point for
understanding the physics of glitches, including the mechanisms of
spin-up and recovery.
The physics of glitches have been made for neutron stars.
The spin-up stage is generally attributed to a transfer of angular
momentum from inner part to the crust of a neutron
star~\citep{Anderson1975}, which could be triggered by starquakes of
an oblate crust~\citep{Ruderman1969, Baym1971} and/or an unpinning
avalanche in the superfluid vortex array~\citep{Warszawski2008,
Melatos2009}.
The spin-up stage lasts for less than tens of seconds, and the
recovery typically lasts for days to weeks~\citep{Wong2001}.

Strangeon star model has been showed to be compatible with
observations of glitches.
Pulsar glitches could be the result of starquakes of solid strangeon
stars~\citep{Zhou2004,PX2008,Zhou2014}.
The detailed modeling about the glitch behaviors compared with
observations has been shown in~\citep[hereafter Paper I]{Lai2018MN}, where the relation
between the recovery coefficients and glitch sizes was found to be
consistent with observations.
The glitch activity of normal radio pulsars~\citep{Lyne2000,Espinoza2011,Fuentes2017} can also be
explained under the framework of starquake of solid strangeon star
model~\citep[Paper II]{WangWH2020}.

The mechanism for recovery stage is more unclear.
The time-scales for the recovery stage are very different from that
of the spin-up stage, suggesting that they involve different physics.
Although the recovery is thought to reflect the restoration of
superfluid-lattice co-rotation by viscous and/or magnetic
forces~\citep{Baym1969, Boynton1972, Lohsen1975}, some aspects of
the glitch recovery have not been explained well, such as the
recovery time-scales~\citep{Wong2001}, the non-single exponential
decays~\citep[e.g.][]{Dodson2002} and the nearly complete recoveries
followed by the secondary spin-ups observed in the Crab
pulsar~\citep{Wong2001}.

In this paper we try to reproduce the exponential recovery of glitches, based on the starquake picture established in Paper I for strangeon stars which is briefly introduced in \S\ref{sec:starquake}.
The recovery process is demonstrated in \S\ref{sec:recovery}, including the recovery model about why and how the recovery would occur.
The time-scale related to the recovery is taken as the viscous time-scale $\tau$ whose values are derived by fitting the recovery data, and it is not necessarily the decay time-scale $\tau_d$ derived by fitting the glitch model.
Although we only consider the single exponential decay which can be reproduced by a uniform viscous flow, this method can be generalized to two or more exponentials by the multi-component of viscous flows.
The magnitude of the recovery is estimated in \S\ref{sec:magnitude} as the result of the recovery of pressure at the cracked places in the equatorial plane, which is consistent with that derived from observations.
Conclusions and discussions are given in \S\ref{sec:conclusion}.

\section{Starquakes of strangeon stars}
\label{sec:starquake}

The recovery follows a starquake, so we will firstly give the picture of starquake process of strangeon stars.
This picture has been described in Paper I, based on the results of~\citet{Baym1971} about the strain in the solid crust of a neutron star, including how the star cracks and how the star reacts to the cracking\footnote{From the elastic deformation theory, the loading of stress field associated with rotation deceleration was derived in~\citet{LuRP2023}.
The results show that the shear stresses near the equator are much larger than that near the poles, and in the rupture the majority of the star volume is characterized by strike-slip faulting under the shear failure, instead of the opening failure used in Paper I.
In fact, although how the starquakes of compact stars deserves to be explored in our future work, it may only affect the details and the qualitative pictures of glitch and recovery in Paper I and this paper would not be changed.}.

\begin{figure}
    \centering
    \begin{minipage}{0.49\columnwidth}
    \centering
    \includegraphics[width=0.9\columnwidth]{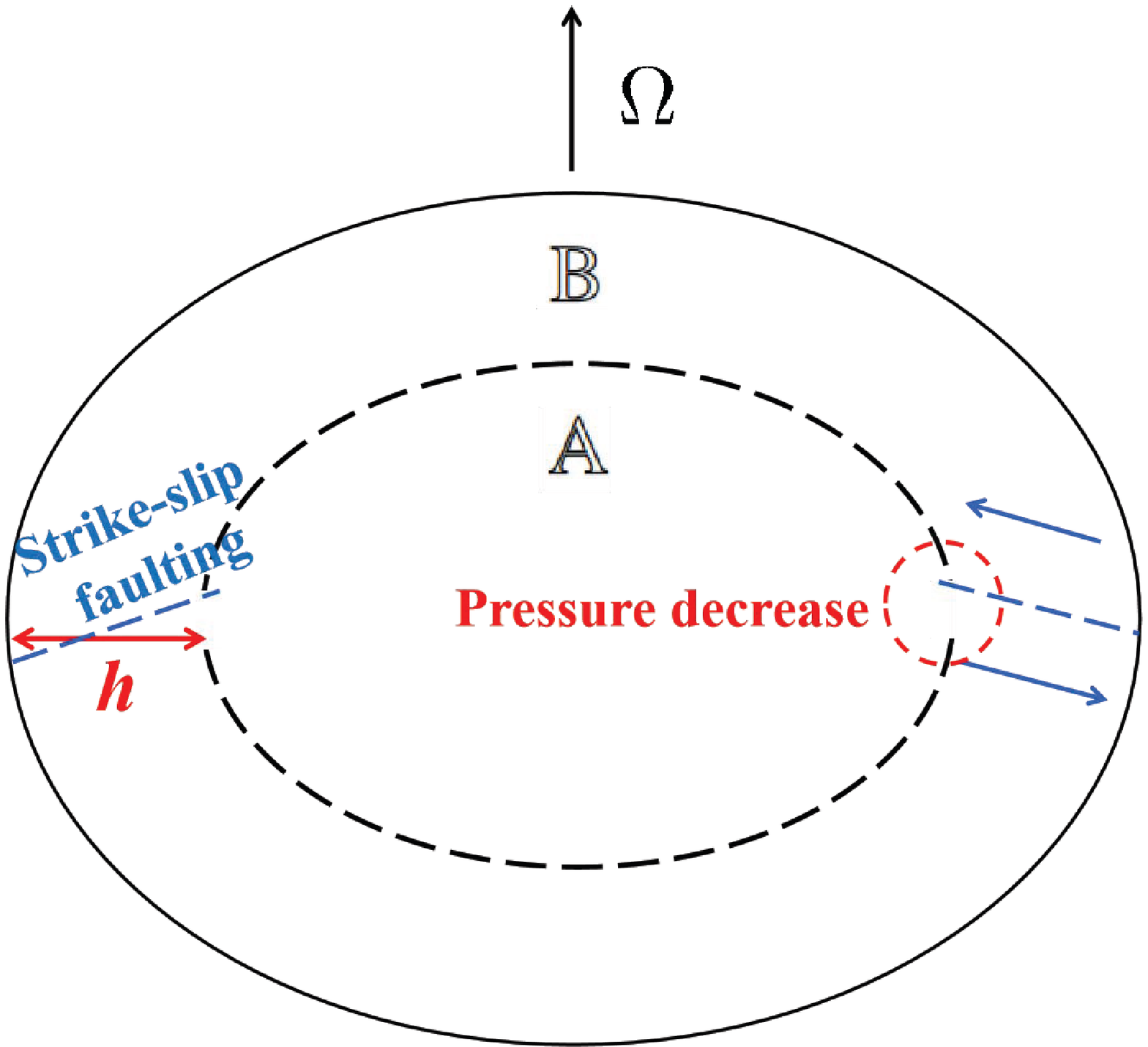}
    \end{minipage}
    \begin{minipage}{0.49\columnwidth}
    \centering
    \includegraphics[width=0.9\columnwidth]{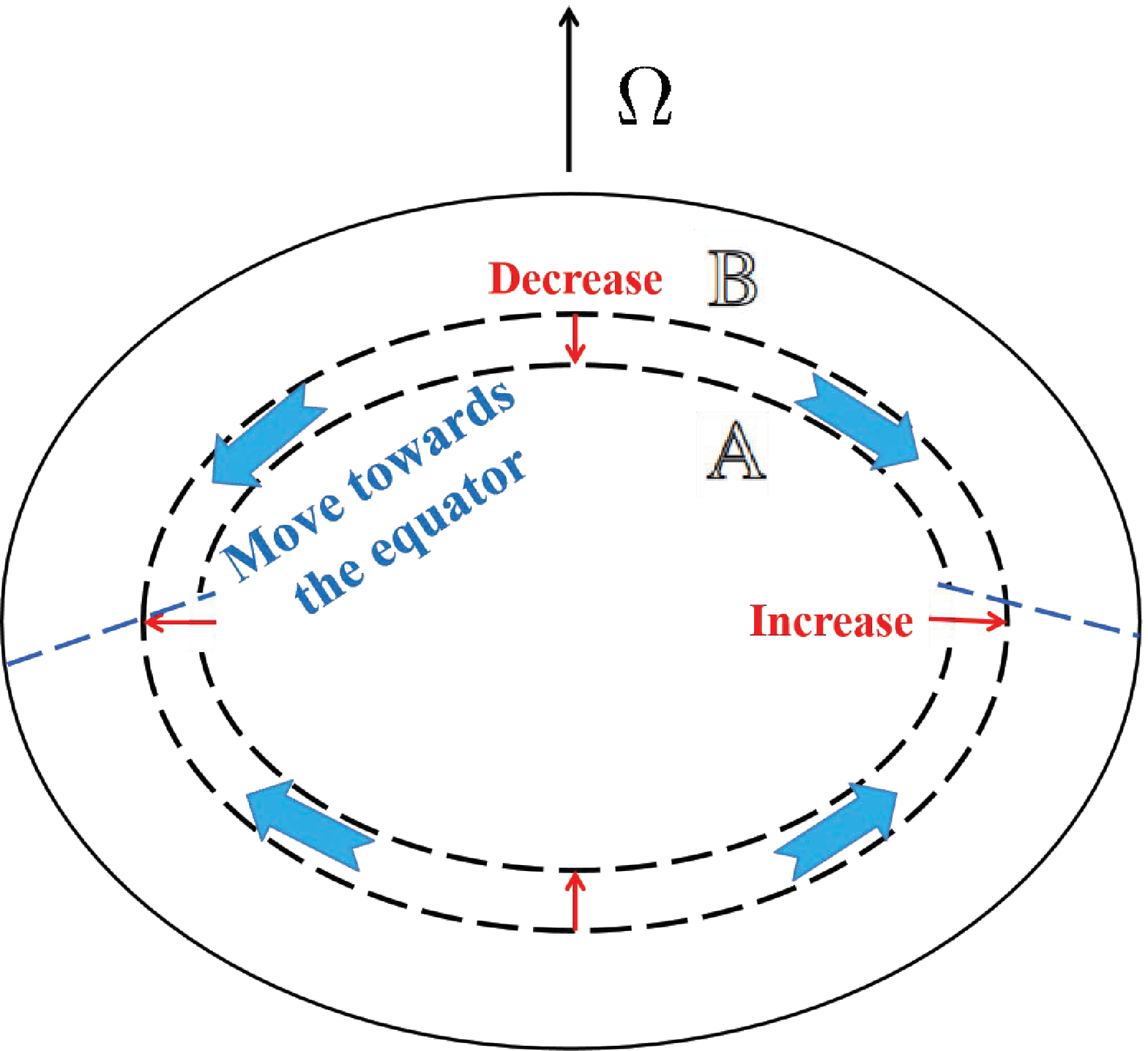}
    \end{minipage}
    \caption{An illustration of the glitch process (the left panel) and recovery process (the right panel).
    Left: The starquake begins with a cracking of the equatorial plane below the surface of the star in the depth $h$ where the critical stress first achieves. After cracking, the layer of the star (labeled by $\mathbb{B}$) breaks along fault lines and moves towards the poles like a plastic flow (Paper I). Right: Because the tangential pressure in the equatorial plane would have been reduced by the inner motion of stellar matter towards the poles, some fragments (in the thin layer between two dashed spheres shown with exaggeration) would flow into the cracking places at the equator, leading to the recovery of the glitch.}
    \label{fig1}
\end{figure}

The starquake process is illustrated in the left panel of Fig.~\ref{fig1}.
Before the starquake, the whole star is an elastic body which is
accumulating the elastic energy.
The starquake begins with a strike-slip faulting in the equatorial plane below
the surface of the star in the depth $h$ where the critical stress
first achieves.
The cracking resulted by the strike-slip faulting then propagates outwards along the equatorial plane.
Being under tension before starquake, after cracking the sphere inside the outer layer (labeled by $\mathbb{A}$) would undergo elastic oscillation.
In the meanwhile, the outer layer of the star (labeled by
$\mathbb{B}$) breaks along fault lines, forming platelets and moving
towards the poles like a plastic flow.
The plastic flow moves tangentially and brings some material from
the equator to the poles.

\subsection{Waiting times}
It is worth emphasizing the following ideas of Paper I about the glitch magnitudes and the waiting times between two glitches.
Although it is a general concept that the required stress develops too slowly
to produce large glitches as often as they are observed to occur in
conventional starquake model of pulsar glitch, our results have shown consistency with the observed values, as explained in the following.
Motivated by the observational fact that glitches with small amplitudes recover almost completely, but those with large amplitudes recover negligible, we introduced a plastic flow (un-recoverable) triggered by oblateness development.
The inner motion of the star during a starquake is not only the change of oblateness, but also a redistribution of matter, both of which would change the moment of inertia of the star.
With the assumption that only the elastic motion, not the plastic flow, would
lead to release of stress during glitches, then we can see that the release of stress is not directly related to glitch sizes $\Delta\nu/\nu$, which means that we cannot predict the time interval $t_{\rm q}$ only from $\Delta\nu/\nu$.
It is consequently reasonable that, the Crab pulsar and the Vela pulsar have nearly the same values of $t_{\rm q}$ although their glitch magnitudes differ by at most three orders of magnitude.
The results shown in Fig.~4 of Paper I are consistent with the observed values.

\subsection{Stress and shear modulus}

The spin-down of a strangeon star will reduce the centrifugal force which causes the accumulation of stress inside the star, and the quake will happen when the critical stress is achieved. The stress distribution of a solid strangeon star during its spin-down has been derived in~\citet{LuRP2023}, provided in Fig.3 of their paper. They found that shear stresses near the equator are much larger than that near the poles. Taken into account the increase of density towards the center, the critical stress would be firstly achieved in the equatorial plane below the surface of the star.

The shear modulus $\mu$ is determined by the interaction between particles inside matter. For the lattice of nuclei of number density $n$, charge $Z$ and lattice constant $a$ interacting via the Coulomb interaction, $\mu\propto Z^2e^2n /a\sim \alpha n^{4/3}$ ($\alpha$ denotes the coupling constant). For the strangeon matter, beside the fact that the number density $n_s$ of strangeons is much larger than $n$ of the neutron star's crust, the strong interaction dominates over the Coulomb interaction by several orders of magnitude, we thus expect that the shear modulus of strangeon matter $\mu_s$ could be 3-5 orders of magnitude larger than that of neutron star's crust $\mu$. Therefore, the shear modulus of strangeon stars $\mu_s$ could be in the range $10^{30}-10^{34} \rm erg/cm^3$~\citep{Xu2003,Zhou2004}.

\subsection{Constraints from gravitational waves}

In Paper I, the deformation due to the starquake in are described by the change of oblateness $\epsilon$, defined as $I=I_0(1+\epsilon)$ where $I$ and $I_0$ are the moments of inertia of deformed and spherical stars respectively. So during the starquake and the recovery stage, the deformation is supposed to be axisymmetric. In this idealized case, the deformation will not be relevant to gravitational waves. 
In fact, however, the deformation will not be purely axisymmetic and will lead to the non-zero ellipticity $\varepsilon$. The large glitches would lead to permanent deformations, so it is interesting to discuss the gravitational waves from the post-glitch phase. 
   
We can estimate the relation between the relative glitch amplitude $\Delta I/I$ and change of ellipticity $\varepsilon$ in the extremely axisymmetic case. 
Because the density of a strangeon star changes not significantly from the center to the surface, we can suppose that the star has a uniform density.
In the extremely axisymmetic case, the equatorial plane changes from a circular (with radius $R$) to an ellipse (with semi-axes $R$ and $a$, $a<R$), then $\Delta \varepsilon=\varepsilon\sim (R-a)/R$.
For the further simplification, we assume that the moment of inertia $I$ is changed by removing the mass $\Delta m$ from the equator to the poles, then $\Delta I/I\sim \Delta m/M\sim R^2(R-a)/R^3\sim (R-a)/R$, which means that $\Delta \varepsilon \sim \Delta I/I$.
This extremely axisymmetic case should put the upper limit to $\Delta \varepsilon$, so $\varepsilon\sim\Delta \varepsilon < \Delta I/I$.
Therefore, observation of gravitational waves might put constraints on our glitch scenario.
For example, the upper limits for $\varepsilon$ of some pulsars with large glitches in the results of the recent LIGO and Virgo data sets~\citep{Abbott_2022} are shown to be larger than $10^{-5}$, consistent with $\varepsilon < \Delta I/I$.

\section{The recovery of glitch}
\label{sec:recovery}



\subsection{The recovery model}
\label{sec:model}

According to Paper I, the recovery of glitch is caused by
the restoring motion of part $\mathbb{A}$, whereas the plastic
motion of part $\mathbb{B}$ would not recover.
It should be noted that, the restoring motion of part $\mathbb{A}$
should not be purely elastic, since it would also be fractured
partially (in the thin layer between two dashed spheres shown with exaggeration in the right panel of Fig.~\ref{fig1}) during the starquake.
Consequently, the time-scale of recovery should be much larger than
that of spin-up.

The inner motion of the stellar matter towards the poles during the glitch would break the
matter between $\mathbb{A}$ and $\mathbb{B}$ into fragments (in the thin layer between two dashed spheres shown with exaggeration in the right panel of Fig.~\ref{fig1}).
In the meanwhile, such inner motion of the stellar matter would reduce
the tangential pressure in the equatorial plane, illustrated in the left panel of Fig.~\ref{fig1}.
Consequently, some fragments would flow into the cracking places and increase the moment of inertia $I$ of
the star, leading to the recovery of glitch, illustrated in the right panel of Fig.~\ref{fig1}.

It is worth noting that, although their real structure should be complex, strangeon stars are supposed to be completely solid in our present work, so there is no phase transition from solid to liquid states. 
After a starquake, the globally rigid body is broken due the motion of fragments, which would be similar to the liquid flow.
In the following, we illustrate the motion of fragments by analogy with the viscous flow, i.e. the ``viscous flow'' in this paper is composed of fragments resulting from the starquake.

In fact, the motion of the viscous flow is essentially driven by the disequilibrium of pressure. The oscillation due to the pressure perturbation is actually a complex process, and it might be an analog of the damped oscillation with time scale depending on its characteristic frequency, which has been discussed in~\citet{Zhou2004}. Here we propose a viscous flow, instead of oscillation, to account for the recovery, whose time scale $\tau$ depends on the properties of the viscous flow, and the possible values of $\tau$ are fitted by the recovery data.

We can estimate the contribution of both parts $\mathbb{A}$ and $\mathbb{B}$ to the star's total moment of inertia $I$.
The density differs not much from the center to the surface of a strangeon star, so we can suppose that the strangeon star has a uniform density.
If the moment of inertia of $\mathbb{B}$, $I_\mathbb{B}$, takes up a fraction $x$ of $I$, i.e. $I_\mathbb{B}=xI$, then $x=1-(1-h/R)^5$.

\subsection{The change of spin-frequency during recovery}
\label{sec:change}

To get a detailed description about this process the increase of $I$
during recovery, we make a simplification that the outflow of matter
in the equatorial plane is equal to the increase of oblateness
$\epsilon$ of ellipsoid $\mathbb{A}$ at an invariant volume $V$ and
density $\rho$.
The moment of inertia $I_\mathbb{B}$ of $\mathbb{B}$ is supposed to
be unchanged during recovery.

Taking the peak of spin-up as our starting point, i.e. the time for
the peak of spin-up is the initial time.
The change of $I_\mathbb{A}$ as time ${\rm d}I_\mathbb{A}/{\rm
d}t\simeq I_\mathbb{A}{\rm d}\epsilon/{\rm d}t$, where
$I_{\mathbb{A}}\propto (1+\epsilon)$.
For an axial symmetric ellipsoid with semi-major axis $a$ and $c$ ($a>c$,
$c$ is along the rotational axis), the oblateness $\epsilon$ is
related to ellipticity $e$ as $\epsilon\simeq e^2/3$ (for
$\epsilon\ll1$).
Then $\epsilon\simeq e^2/3=(1-c^2/a^2)/3=(1-R^6/a^6)/3$, where the
radius $R$ defined as $V=4{\rm \pi} R^3/3$.
Then change rate of oblateness as time can be derived as ${\rm
d}\epsilon/{\rm d}t\simeq (2/R)({\rm d}R/{\rm d}t)\sim (2/R)\upsilon$,
where $\upsilon$ is the velocity of outflow.

The exact motion of outflow is hard to describe, and we make a
further simplification that the outflow is viscous, with the
velocity $\upsilon=\upsilon_0\cdot \exp(-t/\tau)$, where $\upsilon_0$ is the initial
velocity and $\tau$ is the viscous time-scale.
As we mentioned in \S\ref{sec:model}, the viscous flow in this paper refers to the movement of fragments resulting from the starquake. A strangeon star has global rigidity until starquake happens. After the starquake and before the glitch has recovered, some parts of the star would behave like the viscous flow, which would be responsible for the observed properties of the glitch recovery.

Combining with the relation ${\rm d}\epsilon/{\rm d}t= (2/R)\upsilon$, we can get the
change of oblateness of $\mathbb{A}$ as time,
\begin{equation}
\epsilon=\frac{2\upsilon_0\tau}{R}(1-e^{-\frac{t}{\tau}})+\epsilon_0,
\end{equation}
where $\epsilon_0$ is the initial oblateness of $\mathbb{A}$, and
the total increase of oblateness during recovery
$\Delta\epsilon=(\epsilon-\epsilon_0)_{t\rightarrow \infty}=2\upsilon_0\tau/R$.
Then the change of $I_\mathbb{A}$ as time during recovery can be
derived as
\begin{equation}
I_\mathbb{A}=\frac{I_{\mathbb{A}0}}{1+\epsilon_0}[1+\epsilon_0+\Delta\epsilon(1-e^{-\frac{t}{\tau}})],
\end{equation}
where $I_{\mathbb{A}0}$ is the initial value.

The recovery stage stars from spin frequency $\nu_0$, and the
conservation of angular momentum gives
$(I_{\mathbb{A}_0}+I_\mathbb{B})\nu_0=(I_\mathbb{A}+I_\mathbb{B})\nu$,
where $I_\mathbb{B}$ remains unchanged during recovery.
Because $I_\mathbb{B}$ takes up a fraction $x$ of the total moment of
inertia $I$, i.e. $I_\mathbb{B}=xI$ and $I_{\mathbb{A}0}=(1-x)I$, the time evolution of $\nu$ can be derived as
\begin{eqnarray}
\nu(t)
&=&\frac{\nu_0}{x+\frac{1-x}{1+\epsilon_0}\left(1+\epsilon_0+\Delta\epsilon(1-e^{-\frac{t}{\tau}})\right)}\nonumber\\
&=&\frac{\nu_0}{1+\frac{1-x}{1+\epsilon_0}\Delta\epsilon(1-e^{-\frac{t}{\tau}})}\nonumber\\
&=&\frac{\nu_0}{1+\frac{1-x}{1+\epsilon_0}Q\cdot\frac{\Delta\nu_g}{\nu}(1-e^{-\frac{t}{\tau}})}\nonumber\\
&\simeq&\frac{\nu_0}{1+(1-x)Q\cdot\frac{\Delta\nu_g}{\nu}(1-e^{-\frac{t}{\tau}})},
\label{eq:nu}
\end{eqnarray}
where the last approximation is derived from $\epsilon_0\ll1$, and the second to last equation is derived from $\Delta\epsilon=Q\cdot \Delta\nu_g/\nu$, since in our glitch process the recovery of glitch is caused by the restoring motion of part $\mathbb{A}$, and the plastic motion of part $\mathbb{B}$ would not recover (Paper I).
%
%
From the recovery coefficient $Q$ and the relative glitch magnitude $\Delta\nu_g/\nu$, we can get the the time evolution of $\nu(t)$ during recovery.

Before comparing with observations, we would like to show that $\nu(t)$ in Eq.(\ref{eq:nu}) can approach the exponential decay.
The recovery coefficient $Q$ are defined as $Q=\Delta\nu_d/\Delta\nu_g$, where $\Delta\nu_p$ and $\Delta\nu_d$ are respectively the permanent and decay components of the increased frequency $\Delta\nu_g$ of glitches (with $\Delta\nu_g=\Delta\nu_p+\Delta\nu_d$). Then the time evolution of $\nu$ shown in Eq.(\ref{eq:nu}) is $\nu(t)=\nu_0/[1+(1-x)\frac{\Delta\nu_d}{\nu}(1-e^{-\frac{t}{\tau}})]$, which will be approximated to be
\begin{eqnarray}
\nu(t)&\simeq&\nu_0\left[1-(1-x) \frac{\Delta\nu_d}{\nu}\left(1-e^{-\frac{t}{\tau}}\right)\right]\nonumber\\
&\simeq&\nu_0-(1-x)\Delta \nu_d+(1-x)\Delta \nu_d \cdot e^{-\frac{t}{\tau}},
\end{eqnarray}
Then we can see that if $x\ll 1$,
\begin{eqnarray}
\nu(t)&\simeq&\nu_0-\Delta\nu_d+\Delta \nu_d \cdot e^{-\frac{t}{\tau}}\nonumber\\
&=&\nu_{0-}+\Delta \nu_g-\Delta\nu_d+\Delta \nu_d \cdot e^{-\frac{t}{\tau}}\nonumber\\
&=&\nu_{0-}+\Delta \nu_p+\Delta \nu_d \cdot e^{-\frac{t}{\tau}},
\label{eq:nu_approx}
\end{eqnarray}
which differs by only the spin-down terms from the usually used glitch model
\begin{equation}
\nu_{\rm gm}=\nu_{0-}+\Delta\nu_p+(\dot\nu+\Delta\dot\nu_p) t+\Delta\nu_d \cdot e^{-t/\tau_d},
\label{eq:nu_model}
\end{equation}
where $\Delta\dot\nu_p$ are the permanent changes in frequency derivative relative to the pre-glitch values and $\tau_d$ is the decay time-scale.
The reason for this difference is that we do not consider the energy loss in the recovery stage.

\subsection{Comparison with observations}
\label{sec:observation}

The frequency evolution in our recovery model Eq.(\ref{eq:nu}) depends on two parameters, the depth $h$ of the cracking place below the surface ($x=1-(1-h/R)^5$) and the time-scale $\tau$.
The location and shape of the curve are determined by $h$ and $\tau$ respectively.
To get the appropriate values of both parameters, we firstly estimate the value of $\tau$ around $\tau_d$, and then adjust the value of both $h$ and $\tau$ to fit the data.
In addition, because our concern is the recovery stage, we need only to fit the data from $t=0$ (the time of glitch) to $t\sim \tau$.

To compare Eq.(\ref{eq:nu}) with observations, the term $(\dot\nu+\Delta\dot\nu_p) t$ in spin-down model should be subtracted from data.
We choose five glitches, which have observed values of $Q$ and $\Delta\dot\nu_p$~\citep{Yuan2010MN,Dang2020ApJ}, from five pulsars: PSRs J1722-3632, B1800-21, B1823-13, B1838-04 and J1852-0635.

\begin{figure}
    \centering
    \includegraphics[width=0.8\columnwidth]{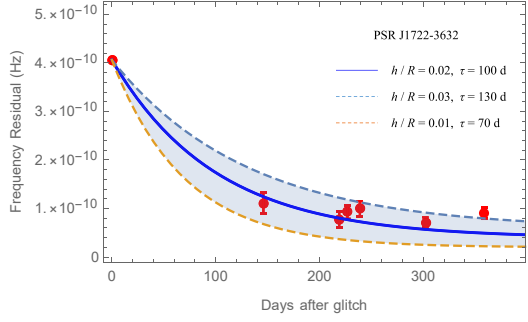}
    \caption{Frequency residual in the recovery for a glitch of PSR J1722-3632 predicted in~Eq.(\ref{eq:nu}), where the permanent jump in frequency has been subtracted to expand the results, is shown in solid line, with $h/R=0.01$ and $\tau=100$ d.
    The red points (except for the first one) are data which have subtracted the spin-down model~\citep{Dang2020ApJ} and also the permanent jump in frequency. The first point is the value at $t=0$ derived from the glitch size and the data before the glitch.
    To see how the predicted curve for frequency residual changes as $h/R$ and $\tau$, we also show the curves with different values of $h/R$ and $\tau$ in dashed lines.
    The blue (upper) and yellow (lower) dashed lines are respectively the results of $h/R=0.03$, $\tau=130$ d and $h/R=0.01$, $\tau=70$ d, which give the range where the appropriate values of $h/R$ and $\tau$ would lie in.}
    \label{fig:nu1}
\end{figure}

Frequency residual in the recovery for a glitch of PSR J1722-3632 predicted in~Eq.(\ref{eq:nu}), where the permanent jump in frequency has been subtracted to expand the results, is shown in solid line in Fig.~\ref{fig:nu1}, with $h/R=0.01$ and $\tau=100$ d.
The red points (except for the first one) are data which have subtracted the spin-down model~\citep{Dang2020ApJ} and also the permanent jump in frequency.
The first point is the value at $t=0$ derived from the glitch size and the data before the glitch.
Solid line is for $h/R=0.02$ and $\tau=100$ d.

To see how the predicted curve for frequency residual changes as $h/R$ and $\tau$, we also show the curves with different values of $h/R$ and $\tau$ in dashed lines.
The blue (upper) and yellow (lower) dashed lines are the results of $h/R=0.03$, $\tau=130$ d and $h/R=0.01$, $\tau=70$ d, respectively.
Because the curve with larger values of $h/R$ and $\tau$ will above the one with smaller values of $h/R$ and $\tau$, the two dashed lines give the range where the appropriate values of $h/R$ and $\tau$ would lie in.


\begin{figure}
    \centering
    \includegraphics[width=0.8\columnwidth]{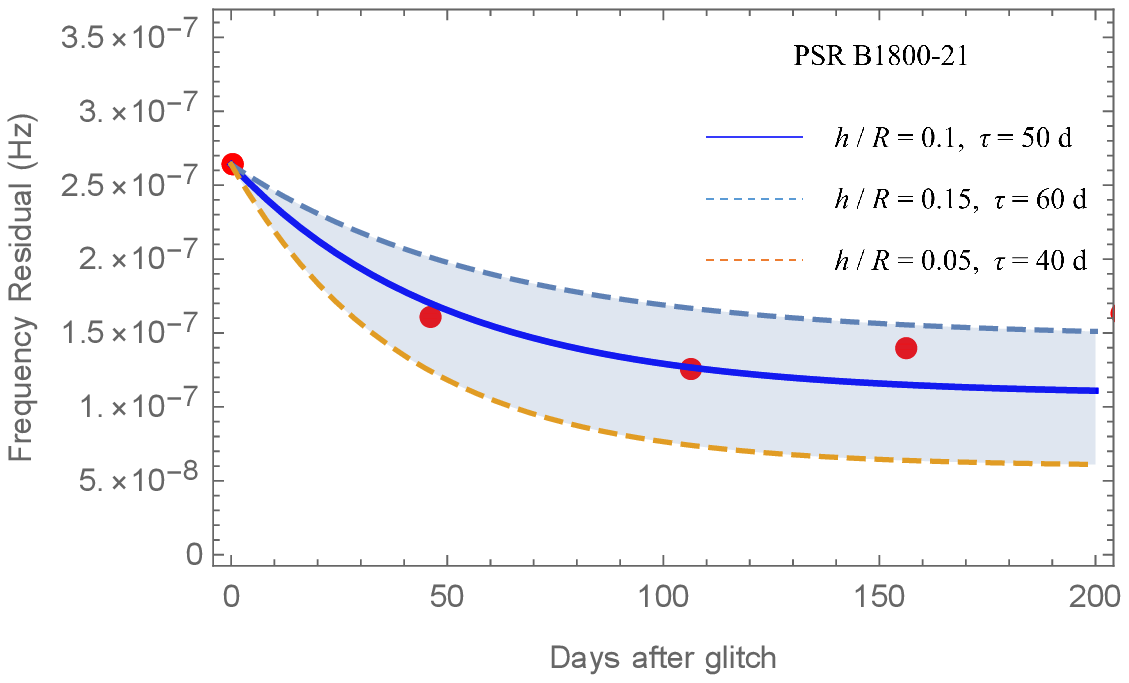}
    \includegraphics[width=0.8\columnwidth]{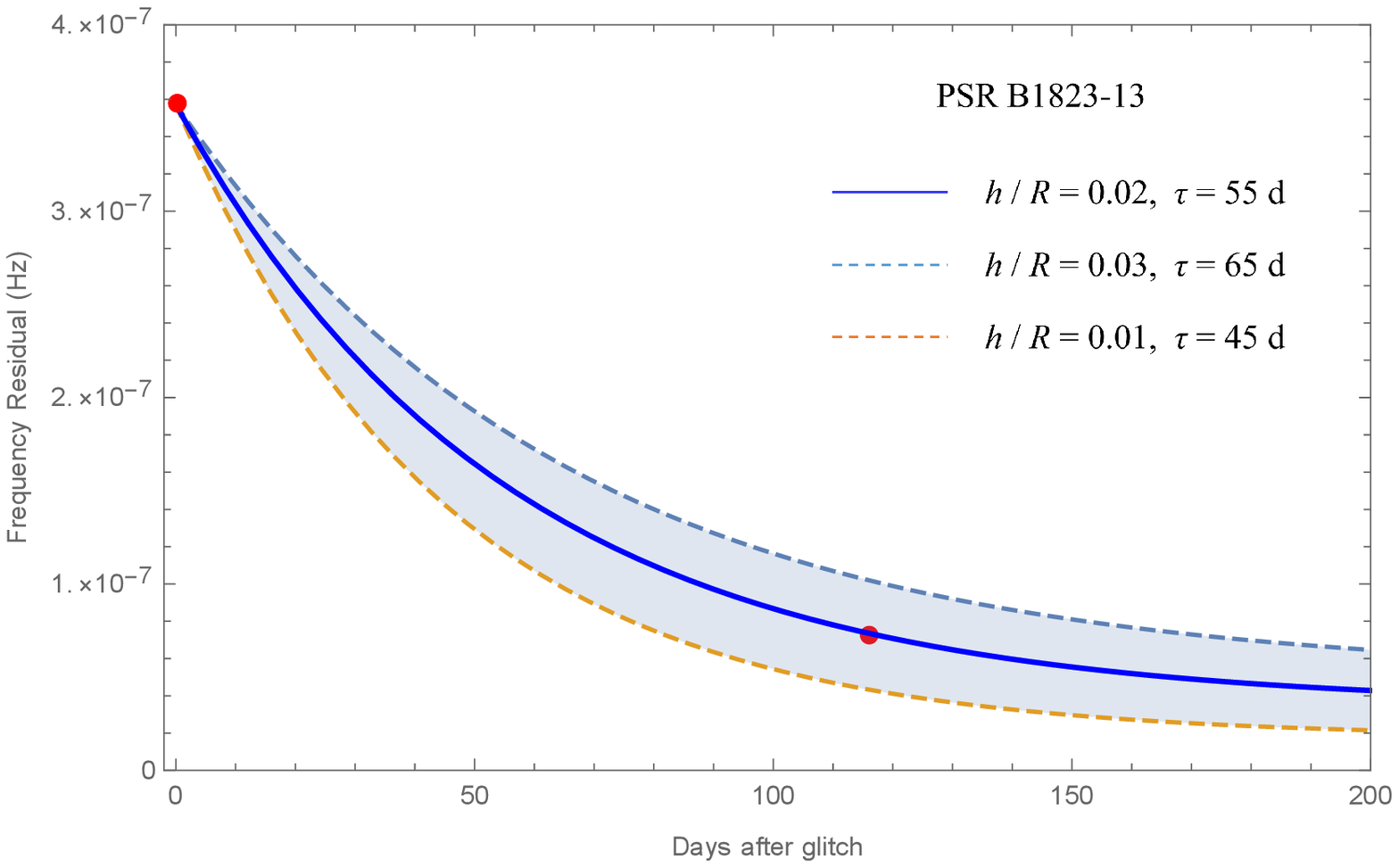}
    \includegraphics[width=0.8\columnwidth]{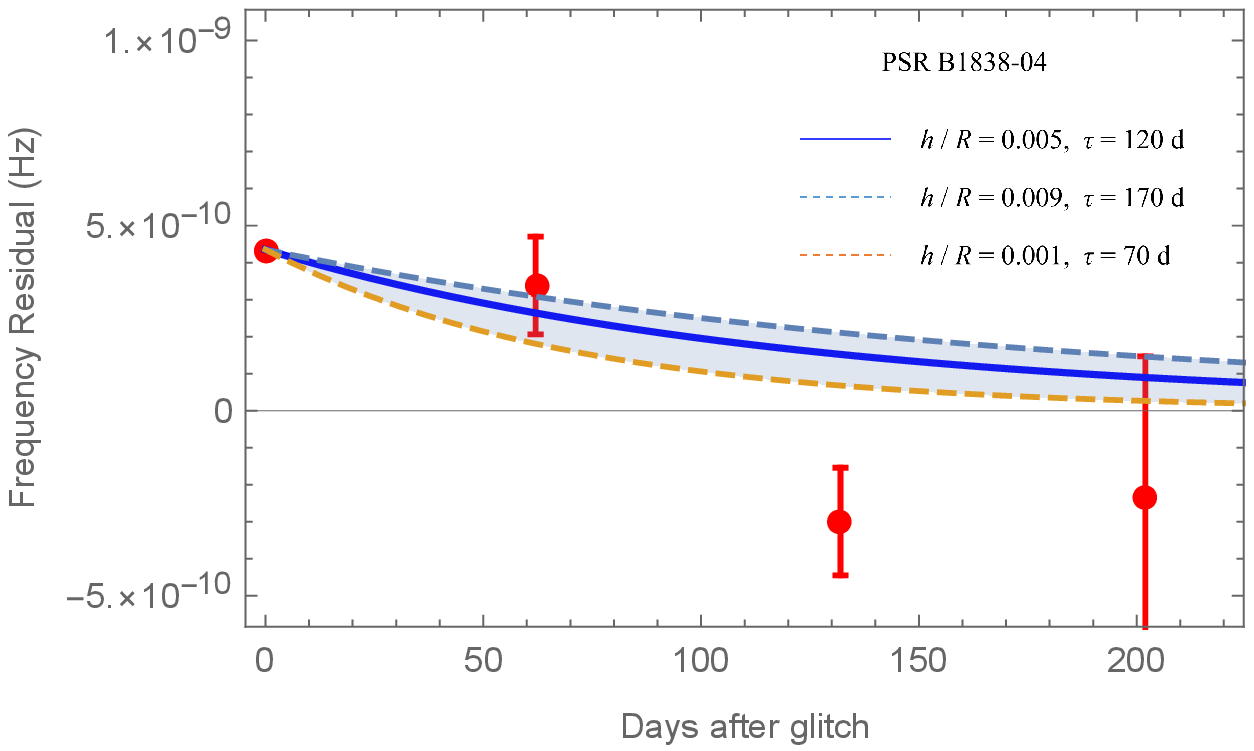}
    \includegraphics[width=0.8\columnwidth]{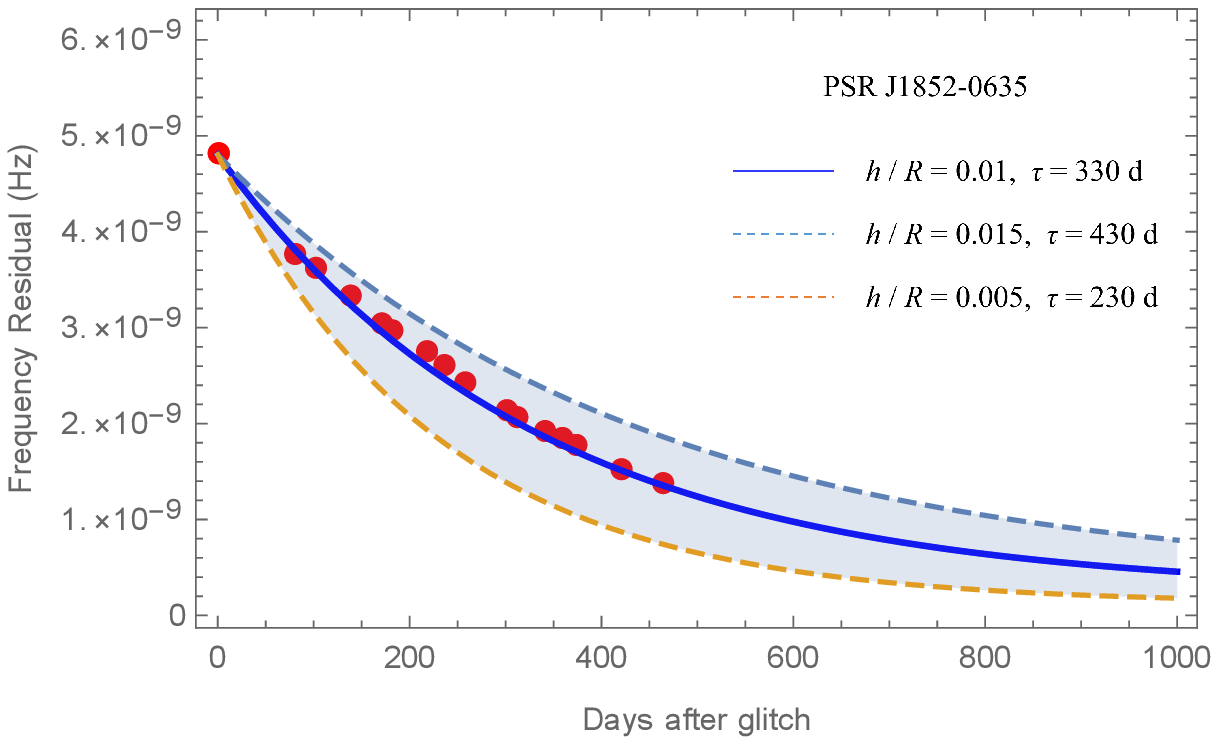}
    \caption{Frequency residuals in the recovery for PSRs B1800-21, B1823-13, B1838-04 and J1852-0635, as the same as~Fig.~\ref{fig:nu1}. The data for PSRs B1800-21, B1823-13, B1838-04 are from~\citet{Yuan2010MN}, and the data for PSR J1852-0635 are from~\citet{Dang2020ApJ}.
    Points and curves are the same as that in Fig.~\ref{fig:nu1}}
    \label{fig:nu4}
\end{figure}

The results for PSRs B1800-21, B1823-13, B1838-04 and J1852-0635 are shown in Fig.~\ref{fig:nu4}, as the same as Fig.~\ref{fig:nu1}.
The data for PSRs B1800-21, B1823-13, B1838-04 are from~\citet{Yuan2010MN}, and the data for PSR J1852-0635 are from~\citet{Dang2020ApJ}.
The values of $h/R$ and $\tau$ for solid and dashed lines are shown in each subplot.

The results of the appropriate values of $h/R$ and $\tau$ shown in solid lines in Fig.~\ref{fig:nu1} and Fig.~\ref{fig:nu4} are listed in Table~\ref{tab:h}.
The values of exponential decay time-scale $\tau_d$ and the relative glitch size $\Delta\nu/\nu$ are also listed~\citep{Yuan2010MN,Dang2020ApJ} in Table~\ref{tab:h}.
We can see that $h/R$ increases with $\Delta\nu/\nu$, which is expected in our glitch model.
The deeper the cracking place is, the larger the glitch size is expected, since the star's spin-up is induced by the move of the fractured part in the outer layer towards the poles (Paper I).

\begin{table}
    \centering
    \caption{The summary in above figures for the five pulsars.
    Columns show respectively the depth $h$ of the cracking place from the surface,
    the viscous time-scale $\tau$ in Eq.(\ref{eq:nu}), the time-scale $\tau_d$ in glitch model of Eq.(\ref{eq:nu_model}),
    the relative glitch size $\Delta\nu/\nu$, and the total increase in radius of the equator during recovery
    $\Delta a$ (see later in~\S\ref{sec:magnitude}). The data of $\tau_d$ and $\Delta\nu/\nu$ are from~\citet{Yuan2010MN} and~\citet{Dang2020ApJ}.}
    \label{tab:h}
    \begin{tabular}{cccccc} 
        \hline
        PSRs & $h/R$ & $\tau\ (d)$ & $\tau_d\ (d)$ & $\Delta\nu_g/\nu\ (10^{-9})$ & $\Delta a$ (cm)\\
        \hline
        B1838-04 & 0.005 & 120 & 80 & 579 & $4.05\times 10^{-5}$\\
        J1852-0635 & 0.01 & 330 & 400 & 1144 & $1.26\times 10^{-3}$\\
        B1823-13 & 0.02 & 55 & 75 & 2416 & $1.81\times 10^{-2}$\\
        J1722-3632 & 0.02 & 100 & 240 & 2702 & $8.11\times 10^{-5}$\\
        B1800-21 & 0.1 & 50 & 120 & 3910 & $1.76\times 10^{-2}$\\
        \hline
    \end{tabular}
\end{table}

It can also be found in Table~\ref{tab:h} that, the viscous time-scale $\tau$ shows no obvious correlation between neither $h/R$, $\tau_d$ nor $\Delta\nu/\nu$.
This may reflect the complexity of the inner motion of the star during glitch and recovery.
The recovery data are from different glitches of different pulsars, and we expect that more recovery data from the same pulsar would reveal how $\tau$ depends on other parameters.

It would be heuristic to learn from the seismology to estimate the recovery time-scale $\tau$.
In the earthquake post-seismic period, the viscous-elastic relaxation is a dominant mechanism controlling the far-field ground deformation. 
In such a mechanism, the characteristic time-scale is estimated from a simple relationship of $\tau=\eta/\mu$~\citep{Rheology2008}, where $\eta$ and $\mu$ are the viscosity and shear modulus, respectively.
The shear modulus of strangeon matter has been estimated to be in the range $10^{30}-10^{34} {\rm erg \cdot cm^{-3}}$~\citep{Xu2003,Zhou2004}, depending on the level of fragmentation~\citep{Precursor2023Zhou}.
Therefore, if $\tau\sim 100 $ d, the viscosity $\eta\sim 10^{37}-10^{41} {\rm erg \cdot s\cdot cm^{-3}}$.
It is interesting to see from Table~\ref{tab:h} that, the two most frequently glitching pulsars among the five, PSRs B1800-21 and B1823-13, have shortest $\tau$ values, i.e. lowest viscosity.
This may reflect the fact that the viscosity $\eta$ depends on the history of starquakes.

\section{The magnitude of recovery}
\label{sec:magnitude}

As demonstrated in \S~\ref{sec:model}, for an axial symmetric ellipsoid with major semi-major axis $a$, the oblateness $\epsilon\simeq(1-R^6/a^6)/3$, where $R$ is star's radius defined as $V=4{\rm \pi} R^3/3$.
The total increase of oblateness during recovery is then $\Delta\epsilon\simeq 2\Delta a/R$, where the total increase of the major semi-major axis is $\Delta a$.
The magnitude of recovery depends on the value of $\Delta a$.
In this section we will demonstrate that, $\Delta a$ predicted in our recovery model is consistent with that derived from observations.

In our glitch picture (Paper I), the increase of oblateness $\Delta\epsilon$ accounts for glitch recovery,
which leads to the relation $\Delta\epsilon=Q\cdot \Delta\nu_g/\nu$.
Therefore, the values of $\Delta a$ can be derived from glitch data, $Q$ and $\Delta\nu_g/\nu$, with $\Delta a\simeq Q\cdot \Delta\nu_g/\nu\cdot R/2$.
If we assume $R=10$ km, then we can get values of $\Delta a$ for the five pulsars we choose in \S\ref{sec:observation}, shown in the last column of Table~\ref{tab:h}.

$\Delta a$ is the change of radius in the equator of the star.
In our glitch picture in Paper I, a glitch is triggered by the starquake which begins with a cracking of the equatorial plane.
The cracking will reduce the pressure of the crack, consequently the matter in the cracking place would be compressed and the radius of the equatorial plane would be reduced.
This contributes to the decrease of moment of inertia that leads to a glitch.
In glitch recovery follows the recovery (i.e. increase) of pressure, via the viscous flow towards the cracked places at equatorial plane.
The role of pressure in the glitch recovery would be similar to that in the post-glacial rebound observed in Iceland on the Earth~\citep{Rebound1991}.

The pressure of the solid strangeon star should be inhomogeneous.
A simplified case is that the inhomogeneous pressure is spherical symmetric, where the tangential pressure $P_{\perp}$ and the radial pressure $P$ are not equal, with $P_{\perp}=(1+\xi)P$.
In this case, the hydrostatic equilibrium equation becomes~\citep{Herrera1997,xty2006}
\begin{equation}
\frac{{\rm d}p}{{\rm d}r}=-\frac{Gm(r)\rho}{r^2}\left(1+\frac{p}{\rho c^2}\right)\left(1+\frac{4\pi r^3p}{m(r)c^2}\right)\left(1-\frac{2Gm(r)}{rc^2}\right)^{-1}+\frac{2\xi p}{r}. \label{eq:dp}
\end{equation}

When the pressure of the crack is reduced by the cracking, the reduce of tangential pressure would be more significant than that of radial pressure.
Therefore, to simplify the problem that how to estimate $\Delta a$, we suppose that only the tangential pressure reduces as the result of cracking.
That means the change of $a$ is the result of the change of $\xi$.
The tangential pressure at first reduces to $\xi^{\prime}$, and recover to $\xi$ (with an increase of $\Delta\xi=\xi-\xi^{\prime}$) as the result of the viscous flow towards the cracked places at equatorial plane.

\begin{figure}
    \includegraphics[width=\columnwidth]{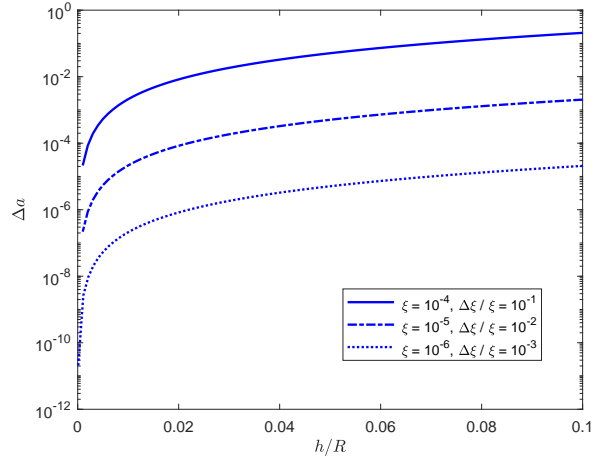}
    \caption{The relation between $\Delta a$ and $h/R$, for different values of $\xi$ and $\Delta\xi/\xi$.
    The values of $\Delta a$ and $h/R$ shown in each row of Table~\ref{tab:h} can be included between the upper (solid) and the lower (dotted) lines.}
    \label{fig:a}
\end{figure}

To derive the value of $\Delta a$, we start from an equation of state, which ensures that $M_{\rm TOV}>2.3 M_\odot$ and $\Lambda(1.4)<800$, and derive the radius $R$ of the star from the hydrostatic equilibrium equation~\ref{eq:dp}.
We then choose a value of $h$ and calculate the difference between $h$ and $h^\prime$, in the cases of $\xi$ and $\xi^\prime$ respectively, and we get $\Delta a =h-h^{\prime}$.
In Fig.~\ref{fig:a} we show the relation between $\Delta a$ and $h/R$, for different values of $\xi$ and $\Delta\xi/\xi$.
We can see that the values of $\Delta a$ and $h/R$ shown in each row of Table~\ref{tab:h} can be included between the upper (solid) and the lower (dotted) lines.

\section{Conclusions and discussions}
\label{sec:conclusion}

In this paper we describe the process of glitch recovery for strangeon stars, based on the starquake picture of glitches established in Paper I.
We give answers to why and how the glitch will recover.
The answer to the first questions is that, the cracking in the equatorial plane splits some stellar matter and reduce the tangential pressure in the equatorial plane, and then the matter in the interface (which separates the plastic and elastic motion during the glitch) would move from high latitudes toward the cracking place in the equatorial plane, leading to the recovery of the glitch.
The answer to the second question is that, because the recovery is the result of the increase of the moment of inertia, we use a viscous outflow of matter in the equatorial plane to calculate the increase of the moment of inertia during recovery, and consequently get the time evolution of spin-frequency $\nu(t)$ during recovery, showing the exponential form.

To compare the theory with observations, we choose five glitches from five pulsars: PSRs J1722-3632, B1800-21, B1823-13, B1838-04 and J1852-0635, which have observed values of $Q$ and $\Delta\dot\nu_p$ to subtract $(\dot\nu+\Delta\dot\nu_p) t$ from data.
This subtraction is needed because we do not consider the energy loss in describing the recovery process, and the spin-down term should be subtracted to from data before the comparison of theory with observations.
We get the ranges for the two parameters, $h/R$ and $\tau$, in $\nu(t)$ by fitting the data, and the results show that $h/R$ increases with the glitch size $\Delta\nu/\nu$.
Such a positive correlation between $h/R$ and $\Delta\nu/\nu$ is consistent with our glitch model, since the deeper the cracking place is, the larger the glitch size is expected in the scenario that the star's spin-up is induced by the move of the fractured part in the outer layer towards the poles (Paper I).
The magnitude of the recovery, estimated as the result of the recovery of pressure at the cracked places in the equatorial plane, is consistent with that derived from observations.

As a first attempt to study the glitch recovery of strangeon stars, this paper gives the above conclusions including both the exponential form and the magnitude of glitch recovery.
Although it is hard to avoid simplifications, the conclusions would not change.
To quantitatively describe the exponential recovery, here we approximate the net result of the matter flow in the interface from high latitudes toward the cracking place to a viscous outflow in the equatorial plane increasing the semi-major axis.
No matter how the redistribution of matter happens after a glitch, it essentially results from the decrease of pressure of the cracking places in the equatorial plane.
The supplement of matter due to the pressure deficit would lead to the increase of the moment of inertia, although in reality it may be achieved by some more complex ways, and the approximation of the viscous flow could be reasonable.
Moreover, we only consider the single exponential decay which can be reproduced by a uniform viscous flow, but this method can be generalized to two or more exponentials by the multi-component of viscous flows.

It is interesting to note that, the surface flow during the starquake would probably change the configuration of magnetic fields and lead to some observable consequences. One observation is from the Vela pulsar. The single-pulse radio observations of a glitch of the Vela pulsar detect sudden changes in the pulse shape that are coincident with the glitch event, indicating that the glitch altered the magnetosphere of the Vela pulsar~\citep{Palfreyman2018}. Another observation is from the Crab pulsar. There are evidences showing that the polarization fraction of the Crab pulsar decreases after a glitch~\citep{Feng2020}, indicating that the properties of magnetic fields may change.

How the starquakes of compact stars happen, and how to describe the inner motion of the star during glitch and its recovery, still remain to be solved.
The loading of stress field associated with rotation deceleration under the elastic deformation theory deserves to be improved in the future.
In addition, the recovery model we propose here involve $Q$ and $\Delta\dot\nu_p$, so we can only use very few data of glitches.
More data of glitches and recoveries are needed to constrain parameters and test theoretical models.


\section*{Acknowledgements}

This work is supported by the National SKA Program of China (2020SKA0120300, 2020SKA0120100), the National Natural Science Foundation of China (Grant Nos. U1831104, 12041304, 42174059), the Outstanding Young and Middle-aged Science and Technology Innovation Teams from Hubei colleges and universities (No. T2021026), and the Young Top-notch Talent Cultivation Program of Hubei Province.

\section*{Data Availability}

The data underlying the work in this paper are available upon reasonable request.












\bsp    
\label{lastpage}
\end{document}